\documentclass[journal]{IEEEtran}
\IEEEoverridecommandlockouts

\usepackage{cite}
\usepackage{amsmath,amssymb,amsfonts}
\usepackage{algorithmic}
\usepackage{graphicx}
\usepackage{textcomp}
\usepackage{xcolor}
\usepackage{hyperref}
\usepackage{setspace} % Add the setspace package

\def\BibTeX{{\rm B\kern-.05em{\sc i\kern-.025em b}\kern-.08em
    T\kern-.1667em\lower.7ex\hbox{E}\kern-.125emX}}

\begin{document}
\title{Model Predictive Control for Integrated Lateral Stability}

\author{%%%% author names
    \IEEEauthorblockN{Jad Yahya}% first author
    , \IEEEauthorblockN{Siddharth Saha}% delete this line if not needed
    , \IEEEauthorblockN{Haoru Xue}% delete this line if not needed
    , \IEEEauthorblockN{Allen Y. Yang}% delete this line if not needed
    % duplicate the line above as many times as needed to list all authors
    \\%%%% author affiliations
    \IEEEauthorblockA{\textit{Department of Mechanical Engineering (University of California, Berkeley), Berkeley, CA}}\\% first affiliation
    \IEEEauthorblockA{\textit{Department of Computer Science (University of California, San Diego) San Diego, CA}}\\
    \IEEEauthorblockA{\textit{The Robotics Institute (Carnegie Mellon University) Pittsburgh, PA}}\\% delete this line if not needed
    \IEEEauthorblockA{\textit{Department of Electrical Engineering and Computer Science (University of California, Berkeley), Berkeley, CA}}\\% first affiliation
    % duplicate the line above as many times as needed to list all affiliations
}

\maketitle

\begin{abstract}
This paper studies the design of a Model Predictive Controller (MPC) for integrated lateral stability, traction/braking control, and rollover prevention of electric vehicles intended for very high speed (VHS) racing applications. We first identify the advantages of a state-of-the-art dynamic model in that it includes rollover prevention into the MPC (a total of 8 states) and also linearizes the tire model prior to solving the MPC problem to save computation time. Then the design of a novel model predictive controller for lateral stability control is proposed aimed for achieving stable control at top speed significantly greater than typical highway speed limits. We have tested the new solution in simulation environments associated with the Indy Autonomous Challenge, where its real-world racing conditions include significant road banking angles, lateral position tracking, and a different suspension model of its Dallara Indy Lights chassis. The results are very promising with a low solver time in Python, as low as 50 Hz, and a lateral error of 30 cm at speeds of 45 m/s. Our open source code is available at: \url{https://github.com/jadyahya/Roll-Yaw-and-Lateral-Velocity-MPC/}.
\end{abstract}


\section{Introduction}
Active stability systems for vehicle control and driver assistance have become extensively used in applications in the past decades \cite{Ataei2019, Dizqah2016, Dai2013}. For lateral stability in particular, active safety control systems are expected to ensure that the vehicle remains in the stable handling region in the face of adversity, such as on slippery surfaces or under aggressive maneuvers, and within the laws of physics. On the contrary, for longitudinal stability control, the systems are expected to ensure that the vehicle wheel spin is regulated so that the braking/acceleration torque is used optimally. Moreover, rollover control systems need to prevent untripped rollovers by regulating lateral load transfer from inner to outer wheels.

In the context of autonomous driving vehicles, various control schemes have been used in the literature to address the issues of stability and position tracking. In \cite{Wang2015}, the \emph{Linear Quadratic Regulator} (LQR) has been studied for roll stability control of buses. LQR was also used in \cite{Li2022}, where it is applied to lateral position tracking. In \cite{Huang2014}, another optimization-based method, H-infinity control, is used for reference yaw rate tracking while overcoming challenges such as backlash-type hysteresis and the constant variance of longitudinal velocity and tire cornering stiffness. Although these controllers seem to perform well, they rely on inherently linear dynamic models, which are not sufficiently accurate when the nonlinear nature of vehicle stability may be pushed to its limit in \emph{very high speed} (VHS) driving or racing applications, typically expecting sustained vehicle speed greater than 60 meters per second (m/s) or equivalently approximately 135 miles per hour (mph). Moreover, they lack predictive capabilities and the ability to account for hard safety constraints. 

Some other studies have considered sliding mode control \cite{Lee2017, Tagne2013, Hatipoglu2003}. However, these works also rely on linear vehicle models or reduced degrees of freedom. In addition, they are known to suffer from the chattering problem especially at the limit of stability, which is common to sliding mode control and could result in poor performance especially at higher speeds.

To overcome these drawbacks, \emph{Model Predictive Control} (MPC) has become widely employed for vehicle stability control \cite{Ataei2019, Dizqah2016, Dai2013}. Its advantages include, firstly, the ability to predict over a finite horizon; secondly, the ability to deal with \emph{Multiple Input Multiple Output} (MIMO) systems; and lastly, the ability to handle constraints. Multiple works in the literature have considered \emph{Nonlinear MPC} (NMPC) models for the aforementioned purposes, but these models are known to cause high computational time \cite{Dizqah2016, Dai2013}. For example, in \cite{Dizqah2016}, the solution had to simplify the NMPC model to only control lateral dynamics without accounting for the coupling effects due to load transfer. In \cite{Dai2013}, the model was reduced to in-plane motion using a bicycle vehicle model. Therefore, an NMPC typically requires compromises in the extent of control by decreasing the number of states controlled.

For the specific applications of VHS racing studied in this paper, the literature on lateral stability control and checkpoint tracking is still very limited. Some work such as \cite{Sukhil2021} involve the development of an adaptive look-ahead pure-pursuit controller . Some other work such as \cite{Ni2017} involve the development of a feedforward and feedback control scheme. However, these approaches either still have overly simplified vehicle dynamics or do not account for some states that would be useful if considered. Finally, in \cite{TUM}, the controller developed is a \emph{Tube Model Predictive Controller} (TMPC). Although the model in this paper is nonlinear, it relies on a rather simple model with only four states for lateral and longitudinal motion. This results in a vehicle dynamics model that relies on states in the plane of motion, with no account for roll dynamics and hence load transfer between the inner and outer wheels during cornering.

This paper is inspired by and extends the work in \cite{Ataei2019}, which aims to develop a linear model predictive controller for slip control, handling improvement, lateral stability control, and rollover prevention of electric vehicles. This is achieved by linearizing the selected vehicle tire model at each solver iteration. In this paper, we will linearize a Pacejka and a Dugoff tire model \cite{Rajamani} (only the Dugoff model will be considered for performance assessment). Additionally, our study will develop a centralized MPC for lateral stability and lateral trajectory tracking of an autonomous racing scenario while accounting for the effects of load shift. This controller then will be deployed to control a real-world Dallara AV-21 vehicle in the Indy Autonomous Challenge. Our team represented by the authors and the others has ranked No. 3 globally in its 2022--2023 season.

The Python implementation of the solution is open sourced at: {\it https://github.com/jadyahya/Roll-Yaw-and-Lateral-Velocity-MPC/}.

\section{Vehicle Dynamic Model}
The dynamic model used in this paper both for the general electric vehicle and for our experiments on the Dallara AV-21 chassis is an in-plane model with roll dynamics \cite{Ataei2019}. To develop this model, the tire model was initially linearized about an equilibrium point into the following form:
\begin{equation}
\left\{
\begin{matrix}
f_{yi} = \bar{f}_{yi} + \tilde{c}_{\alpha_i}(\alpha_i-\bar{\alpha}_i) \\
f_{xi} = \frac{Q_i}{R_w}
\end{matrix}
\right.
\quad \text{for } i=1 \text{ to } 4,
\label{eq:linear-tire-model}
\end{equation}
where \(f_{y i}\) is the tire force, \(\bar{f}_{y i}\) and \(\tilde{c}_{\alpha_i}\) represent the lateral tire force and the cornering coefficient at the side slip angle of the operating point, respectively, and the front-left, front-right, rear-left, and rear-right wheels are labeled as wheel number 1–4, respectively. \(f_{x i} \) is the longitudinal force on the tires, \(Q_{i}\) is the torque on each wheel and \(R_w\) is the wheel's effective radius.
But, from vehicle dynamics, we have:
\begin{equation}
    \alpha_i=\delta_i-\frac{v+a_i r}{u},
    \label{eq:alpha_i}
\end{equation}
where \(\delta_{i}\) is the steering angle for tire \(i\), $u$, $v$, and $r$ are the longitudinal velocity, lateral velocity, and yaw rate and \(\varphi\) is the roll angle. Also, \(l_{f}\) and \(l_{r}\) are the horizontal distances of the \emph{center of gravity} (CoG) to the front and rear axles, respectively. Ataei et al. define: 
\begin{equation}
    a_i= \begin{cases}l_f, & i=1,2 \\ -l_r, & i=3,4\end{cases}.
    \label{eq:a_i}
\end{equation}

Using \eqref{eq:linear-tire-model}, \eqref{eq:alpha_i}, and \eqref{eq:a_i}, they develop the following equation
\begin{equation}
    f_i = B_{1 i} X_b + B_{2 i} W_i + D_{1 i},
    \label{eq:tireforces}
\end{equation}

where
\begin{equation}
    f_i=\left[f_{x i}, f_{y i}\right]^T,
    \label{eq:tireforcedef}
\end{equation}
\begin{equation}
    W_i=\left[Q_i, \delta_i\right]^T,
    \label{eq:inputdef}
\end{equation}
\begin{equation}
    B_{1 i}=\left[\begin{array}{cccc}
0 & 0 & 0 & 0 \\
-\frac{\tilde{\alpha}_{\alpha_i}}{u} & -\frac{a_i \tilde{c}_{\alpha_i}}{u} & 0 & 0
\end{array}\right],B_{2 i}=\left[\begin{array}{cc}
\frac{1}{R_w} & 0 \\
0 & \tilde{c}_{\alpha_i}
\end{array}\right],
\label{eq:b1ib2i}
\end{equation}
\begin{equation}
    D_{1 i}=\left[\begin{array}{c}
0 \\
\bar{f}_{y i}-\tilde{c}_{\alpha_i} \bar{\alpha}_i
\end{array}\right],
\label{eq:d1i}
\end{equation}
\begin{equation}
    X_b=[v,  r,  \varphi,  \dot{\varphi}]^T.
    \label{eq:xb}
\end{equation}

They further define a local actuator reconfiguration matrix for each wheel. It is a diagonal matrix that takes binary entries for its diagonal entries to indicate whether the actuator is available  for vehicle control or not and is given by: 
\begin{equation}
    T_{w i}=\left[\begin{array}{cc}
t_{\mathrm{Q} i} & 0 \\
0 & t_{\delta i}
\end{array}\right].
\label{eq:actconfig}
\end{equation}

In addition, a mapping matrix \(L_{w i}\) is defined for the mapping from the local tire forces to vehicle corner forces (in the direction of the axes attached to the vehicle's center of gravity):
\begin{equation}
    L_{w i}=\left[\begin{array}{cc}
\cos \delta_i & -\sin \delta_i \\
\sin \delta_i & \cos \delta_i
\end{array}\right].
\label{eq:lw}
\end{equation}
The corner forces at the \(i_{th}\) wheel were defined as
\begin{equation}
    F_{c i}=\left[F_{x i}, F_{y i}\right]^T.
    \label{eq:cornerforcesdef}
\end{equation}
The control input utilized was given by
\begin{equation}
\begin{aligned}
    &\delta f_i=B_{2 i} T_{w i} U_i, \\
    &\delta f_i=\left[\delta f_{x i}, \delta f_{y i}\right]^T, \\
    &U_i=\left[\Delta Q_i, \Delta \delta_i\right]^T.
\end{aligned}
\label{eq:controlinput}
\end{equation}

Utilizing \eqref{eq:tireforces}, \eqref{eq:actconfig}, \eqref{eq:lw}, \eqref{eq:cornerforcesdef}, and \eqref{eq:controlinput}, the following equation was developed:
\begin{equation}
    F_{c i}=L_{w i}\left(B_{1 i} X_b+B_{2 i} W_i+D_{1 i}+B_{2 i} T_{w i} U_i\right).
    \label{eq:cornerforces}
\end{equation}
Combining the equations for all 4 tires into one equation yields:
\begin{equation}
    F_c=L_w\left(B_1 X_b+B_2 W+D_1+B_2 T_w U\right),
    \label{eq:cornerforcesconcat}
\end{equation}
where:
\begin{equation*}
    \begin{aligned}
        &F_c=\left[F_{c 1}^T, F_{c 2}^T, F_{c 3}^T, F_{c 4}^T\right]^T,\\
        &U=\left[U_1^T, U_2^T, U_3^T, U_4^T\right]^T,\\
        &W=\left[W_1^T, W_2^T, W_3^T, W_4^T\right]^T, \\
        &L_w=\operatorname{blockdiag}\left(L_{w 1}, L_{w 2}, L_{w 3}, L_{w 4}\right),\\
        &B_1=\left[B_{11}^T, B_{12}^T, B_{13}^T, B_{14}^T\right]^T, \\
        &B_2=\operatorname{blockdiag}\left(B_{21}, B_{22}, B_{23}, B_{24}\right), \\
        &D_1=\left[D_{11}^T, D_{12}^T, D_{13}^T, D_{14}^T\right]^T,\\
        &T_w=\operatorname{blockdiag}\left(T_{w 1}, T_{w 2}, T_{w 3}, T_{w 4}\right). \\
    \end{aligned}
\end{equation*}

Moreover, the corner forces were translated into forces at the vehicle's center of gravity. The force vector on the center of gravity, denoted by \(F\), can be expressed as
\begin{equation}
    F=\left[
F_X, F_Y, M_Z
\right]^T.
\end{equation}

Define \(L_c\) the mapping matrix from corner forces to CoG forces as 
\begin{equation}
    L_c=\left[\begin{array}{cccccccc}
1 & 0 & 1 & 0 & 1 & 0 & 1 & 0 \\
0 & 1 & 0 & 1 & 0 & 1 & 0 & 1 \\
-t_f / 2 & l_f & t_f / 2 & l_f & -t_r / 2 & -l_r & t_r / 2 & -l_r
\end{array}\right],
\end{equation}
where \(t_f\) and \(t_r\) are the vehicle's front and rear track widths. Thus, the following relation was established:
\begin{equation}
    F=L_c F_c.
    \label{eq:cornertocg}
\end{equation}

Figure \ref{fig:diagram-different-forces} below shows the different kinds of forces on the vehicle. In blue are the tire forces, in red are the corner forces, and in the direction of the red vectors are the forces on the vehicle's CoG.
\begin{figure}[h!]
\centering
\includegraphics[width=80mm]{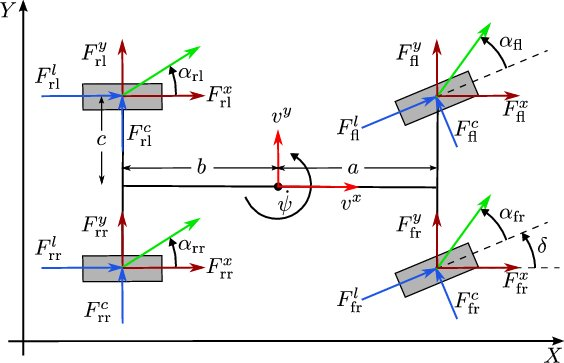}
\caption{Diagram Visualizing the Different Kinds of Forces\cite{Zanon2014}.}
\label{fig:diagram-different-forces}
\end{figure}

From here on, the differences between the general electric vehicle model and the Dallara AV-21 model become pronounced. The different models will be discussed in the below subsections.

\subsection{General Electric Vehicle Model}

The vehicle body dynamics for the general electric vehicle were derived from Newton's equations of motion.
\begin{equation}
    \dot{X}_b=A_F X_b+B_F F,
    \label{eq:ssbutFinput}
\end{equation}
where
\begin{equation}
        A_F=\left[\begin{array}{cccc}
0 & -u & \frac{m_s h_s\left(k_{\varphi}-m_s g h_s\right)}{\left(-m_s^2 h_s^2+m I_{x x}\right)} & \frac{m_s h_s c_{\varphi}}{\left(-m_s^2 h_s^2+m I_{x x}\right)} \\
0 & 0 & 0 & 0 \\
0 & 0 & 0 & 1 \\
0 & 0 & \frac{-m\left(k_{\varphi}-m_s g h_s\right)}{\left(-m_s^2 h_s^2+m I_{x x}\right)} & \frac{-m c_{\varphi}}{\left(-m_s^2 h_s^2+m I_{x x}\right)}
\end{array}\right]\\,
\end{equation}
\begin{equation}
    B_F=\left[\begin{array}{ccc}
0 & \frac{I_{x x}}{\left(-m_s^2 h_s^2+m I_{x x}\right)} & 0 \\
0 & 0 & \frac{1}{I_{z z}} \\
0 & 0 & 0 \\
0 & \frac{m_s h_s}{\left(-m_s^2 h_s^2+m I_{x x}\right)} & 0
\end{array}\right].
\end{equation}
Here, \(m\) and \(m_s\) are the total mass and the sprung mass, respectively. \(I_{xx}\) and \(I_{zz}\) are the roll and yaw moments of inertia, g is the gravitational acceleration, \(h_s\) is the distance between the vehicle's CoG and its roll center, and \(k_\varphi\) and \(c_\varphi\) represent the effective torsional stiffness and torsional damping in the roll direction, respectively.
Joining \eqref{eq:cornertocg} and \eqref{eq:ssbutFinput} yields
\begin{equation}
\begin{aligned}
    \dot{X}_b=\left(A_F+B_F L_c L_w B_1\right) X_b
    +\left(B_F L_c L_w B_2\right) W\\ + \left(B_F L_c L_w B_2 T_w\right) U+B_F L_c L_w D_1.
\end{aligned}
\end{equation}
The wheel dynamics were derived using
\begin{equation}
    I_w \dot{\omega}_i=Q_i+\Delta Q_i-R_w f_{x i},
\end{equation}
where \(I_w\) and \(\omega_i\) are the rotational moment of inertia and the rotational speed. Defining \(X_w=\left[\begin{array}{llll}
\omega_1 & \omega_2 & \omega_3 & \omega_4
\end{array}\right]^T\), the state-space equation for all four wheels can be written as
\begin{equation}
    \dot{X}_w=A_w X_w+E_w W+B_w U+D_w.
\end{equation}
Finally, the full standard state-space form was expressed as
\begin{equation}
    \begin{aligned}
        &\dot{X}=A X+E W+B U+D, \\
        \text{where}\\
        &X=\left[v, r, \varphi, \dot{\varphi}, \omega_1, \omega_2, \omega_3, \omega_4\right]^T, \\
        &U=\left[\Delta Q_1, \Delta \delta_1, \Delta Q_2, \Delta \delta_2, \Delta Q_3, \Delta \delta_3, \Delta Q_4, \Delta \delta_4\right]^T, \\
        &A=\operatorname{blockdiag}\left(A_b, A_w\right), \\
        &E=\left[E_b^T, E_w^T\right]^T, \\
        &B=\left[
        B_b^T, B_w^T \right]^T, \\
        &D=\left[D_b^T, D_w^T\right]^T, \\
        &A_b=A_F+B_F L_c L_w B_1, \\
        &E_b=B_F L_c L_w B_2, \\
        &B_b=B_F L_c L_w B_2 T_w, \\
        &D_b=B_F L_c L_w D_1.
    \end{aligned}
\end{equation}

\subsection{VHS Vehicle Model}
In this subsection, we further derive a special vehicle dynamic model for 
VHS applications, which will later be applied to the Dallara AV-21 chassis used in the Indy Autonomous Challenge in Section \ref{sec:parameters-dallara}.

The model will be modified as show below, albeit still derived from Newton's equations of motion:
\begin{equation}
    \dot{X_b}= A_F X_b + B_F F +C_\phi \phi_r,
\end{equation}
where
{
\scriptsize
\begin{center}

\begin{equation}
    \begin{aligned}
    A_F = \begin{bmatrix}
    0 & 1 & 0 & 0 & 0 \\
    0 & 0 & -u & \frac{- g h_s^2 m m_s + 0.5 * h_s k_s l_ s^2 m_s}{m(I_{xx} + h_s^2 m - h_s^2 m_s)} & \frac{b_s h_s l_s^2 m_s}{(m(I_{xx} + h_s^2 m - h_s^2 m_s))} \\
    0 & 0 & 0 & 0 & 0 \\
    0 & 0 & 0 & 0 & 1 \\
    0 & 0 & 0 & \frac{g h_s m^2 - 0.5*k_s * l_s^2 m}{m(I_{xx} + h_s^2 m - h_s^2 m_s)} & -\frac{b_s l_s^2 m}{m(I_{xx} + h_s^2 m - h_s^2 m_s)}
    \end{bmatrix},
    \\ B_F = 
    \begin{bmatrix}
        0 & 0 &0\\
        0 & \frac{I_xx+m h_s^2}{m(I_{XX} +h_s^2 m -h_s^2*m_s} & 0\\
        0 & 0 & \frac{1}{I_z}\\
        0 & 0 & 0\\
        0 & \frac{m h_s}{2m(I_{xx} +h_s^2 m -h_s^2*m_s} & 0
    \end{bmatrix},
    C_\phi =  
    \begin{bmatrix}
        0 \\
        \frac{g m^2 h_s^2 + gI_{xx}m}{m(I_{xx} +h_s^2 m -h_s^2*m_s)} \\
        0 \\
        0 \\
        \frac{ -g h_s m^2}{m(I_{xx} +h_s^2 m -h_s^2*m_s)}
    \end{bmatrix},
    \end{aligned}
\end{equation}
\end{center}
}
and
\begin{equation*}
    X_b = \begin{bmatrix}
        y&
        v_y&
        r&
        \varphi
        &\dot{\varphi}
    \end{bmatrix}^T,
\end{equation*}
where \(\varphi_r\) is the road banking angle, \(k_s\) is the spring stiffness and \(b_s\) is the damping of the damper.

Additionally, in this case, the changes implemented are adding the lateral position as a state in order to ensure the racing line is tracked for the autonomous racing car. Moreover, the banking angle is added to the model since oval tracks tend to have significant banking angles. This entailed a small angle assumption on \(\varphi - \phi_r\) which is common for vehicle dynamics applications. Moreover, the suspension model was changed from including the entire vehicle to a spring-damper model. We are aware that the stiffness and damping values are different for the front and rear suspension, but for the purposes of this application we test the effect of equating them.

Lastly, we do not consider tire longitudinal slip control and hence do not include the wheel rotational speeds as states. The reasoning behind this is that the vehicles used in the Indy Autonomous Challenge do not hit their maximum acceleration and hence should not need this kind of control (especially at the cost of adding 4 states to the MPC). Moreover, in the case that is to happen, the team has already developed an ABS system to ensure no excessive tire slip occurs. Hence, the state space model in this case is given by
\begin{equation}
    \begin{aligned}
        \dot{X}_b=\left(A_F+B_F L_c L_w B_1\right) X_b+\left(B_F L_c L_w B_2\right) W
        \\+\left(B_F L_c L_w B_2 T_w\right) U+B_F L_c L_w D_1 +C_\phi \phi_r.
    \end{aligned}
\end{equation}

\section{MPC Design and Algorithms}

In formulating the MPC problem, we use two different approaches for the different cases. This will be elaborated on in the following subsections.

\subsection{General Electric Vehicle Model}

After developing the vehicle dynamic model, a controller is  recreated from the work in [1] that considers integrated handling improvement, vehicle lateral stability, rollover prevention, and slip control in braking and traction for electric vehicles.
First, they consider yaw rate tracking which requires a desired yaw rate. That is considered to be the response of a linear bicycle model given by
\begin{equation}
    r_b=\frac{u}{l+k_{u s_{-} d} u^2} \delta_d,
\end{equation}
where l is the vehicle's wheelbase, \(\delta_d\) is the driver's steering angle,and \(k_{u s_{-} d}\) is the desired under-steer coefficient for the vehicle. However, by employing the maximum lateral force on each axle, the below relation was derived
\begin{equation}
    r_{\max }=\frac{\mu_y g}{u},
\end{equation}
where \(\mu_y\) is the lateral friction coefficient. So, the desired yaw rate is defined by
\begin{equation}
    r_d=\operatorname{sign}\left(\delta_d\right) \times \min \left(\left|r_b\right|, r_{\max }\right).
\end{equation}
Next, the lateral stability constraint was considered. To do so, the side-slip angle of the rear tire was limited as 
\begin{equation}
    \left|\frac{-v_y}{u}+\frac{l_r}{u} r\right|<\alpha_{r_{-} \max },
\end{equation}
where \(\alpha_{r_{-} \max}\) is the maximum allowable side-slip angle for the rear tire.
Third, we resort to the rollover index for rollover prevention. Namely
\begin{equation}
    R I=C_1 \varphi+C_2 \dot{\varphi},
\end{equation}
where 
\small
\begin{equation}
    \begin{gathered}
        C_1=\frac{2}{m g T}\left(k_{\varphi}\left(1+\frac{m_s h_R+m_u h_u}{m_s h_s}\right)-\left(m_s h_R+m_u h_u\right) g\right), \\
        C_2=\frac{2 c_{\varphi}}{m g T}\left(1+\frac{m_s h_R+m_u h_u}{m_s h_s}\right).
    \end{gathered}
\end{equation}
\normalsize
Here, \(m_u\) is the unsprung mass, \(h_u\) is the CoG height of the unsprung mass, and \(h_R\) is the distance of the  roll center to the ground.

Finally, for slip control for traction and braking, we resort to the following expression [1]
\begin{equation}
    \omega_{i_{-} c}=\frac{u}{R_w} \pm \lambda_{\max } \max \left(\frac{u}{R_w}, \omega_i\right),
\end{equation}
where \(\lambda_{max}\) is the maximum allowable slip ratio.The set of state constraints turns out to be
\begin{equation}
    \begin{aligned}
        -RI_{max}\leq RI \leq RI_{max},\\
        -r_{max}\leq r \leq r_{max},\\
        min\{\omega_{i{-}c}\} \leq \omega_i \leq max\{\omega_{i{-}c}\}, \\
            -\alpha_{r_{-}\max} \leq \frac{v_y}{u}+\frac{l_r}{u} r\leq \alpha_{r_{-} \max} .
    \end{aligned}
    \label{eq: stateconstraints}
\end{equation}
As for actuator constraints, the ones considered in this problem relate to the maximum possible torque generated, the maximum tire force capacity according to the friction between tires and the road, and the maximum steering angle.
\begin{equation}
    \begin{aligned}
        Q_i^{min}-Q_i(0) \leq\Delta Q_i^k\leq Q_i^{max}-Q_i(0)\\
        -f_{xi}^p(0)-Q_i(0) \leq\Delta Q_i^k\leq f_{xi}^P(0) -Q_i(0)\\
        \delta_i^{min}-\delta_i(0) \leq\Delta \delta_i^k\leq \delta_i^{max}-\delta_i(0)\\
    \end{aligned}
    \label{eq: inputconstraints}
\end{equation}
where \(Q_i(0)\) is the driver command at the beginning of the corresponding horizon, \(\delta_i(0)\) is the driver steering command at the beginning of the corresponding horizon, and \(f_{xi}^p(0) = \mu_x f_{z i}(0) \sqrt{1-\left(\frac{f_{y i}(0)}{\mu_y f_{z i}(0)}\right)^2}\). Also, \(f_{z i}(0) \text { and } f_{y i}(0)\) are the vertical and lateral forces at the beginning of the horizon. 

With these constraints defined as shown in \eqref{eq: stateconstraints} and \eqref{eq: inputconstraints}, the \emph{constrained finite time optimal control} (CFTOC) problem 
 is formulated in Pyomo to track a desired state vector given by 
\begin{equation}
    \bar{X_d} = \begin{array}{cccccccc}
         [0, & r_d, &0, & 0, & \frac{u}{R_w}, &\frac{u}{R_w}, & \frac{u}{R_w}, &\frac{u}{R_w}]
    \end{array}
\end{equation}.

\subsection{VHS Vehicle Model}
The MPC design in this case should maintain lateral stability and lateral race line tracking. Hence, it would use the same actuator constraints as those in \eqref{eq: inputconstraints}. However, from \eqref{eq: stateconstraints}, we only need to consider the second and fourth equations. Also, the model assumes the front wheels are not driven and the rear wheels cannot steer. This last point is reflected in the choice of entries in \eqref{eq:actconfig}. 

Following identifying the constraints, we know we want to track the following set of states 
\begin{equation}
    \bar{X}_d = [y_d, 0, r_d, 0, 0].
\end{equation}
The problem was formulated as a CFTOC in Pyomo.\footnote{\url{http://www.pyomo.org/}. A Python-based, open-source optimization modeling language.} Moreover, to make this feasible for real-time implementation, we have developed it as a Quadratic Program with substitution in CasADi\footnote{\url{https://web.casadi.org/}. An open-source software tool for numerical optimization.} and solving it using the solver 'osqp' (Operator Splitting Quadratic Program)\footnote{\url{https://osqp.org/}.}, which is considered one of the fastest and most reliable optimization solvers for the purposes of MPC design..

The desired steering angle for tracking the racing line is developed by utilizing checkpoints provided by a path planner. The latter provides a desired coordinate to track \((x_d,y_d)\). This can then be utilized to determine the desired heading angle as \(\psi_d = \operatorname{atan}(\frac{y_d}{x_d})\). Therefore, the desired steering angle is determined as \(\delta_d = \psi_d+\frac{v_y}{u}\).

\section{Parameters}
\subsection{General Electric Vehicle Model}
While we were  working on gaining access to a high-fidelity simulator we  relied on a model with full-state feedback and perfect model knowledge to understand the controllers performance as well as debug it. For a general electric vehicle, the parameters used are defined in Table \ref{tab:ev_parameters}.
\begin{table}[htbp]
\caption{Vehicle Parameters for General Electric Vehicle.}
\label{tab:ev_parameters}
\begin{center}
\begin{tabular}{l|l|l|l}
\hline
Parameter & Value &Parameter&Value\\
\hline
\(m_s (kg)\) & 1590 &
\(m_u (kg)\) & 270 \\
\(t_f = t_r (m)\) & 1.575 &
\(h_cg (m)\) & 0.72 \\
\(l_f (m)\) & 1.18 &
\(l_r (m)\) & 1.77 \\
\(I_{xx} (kg m^2)\) & 894.4 &
\(I_{zz} (kg m^2)\) & 2687.1 \\
\(h_s (m)\) & 0.57 &
\(h_u (m)\) & 0.2 \\
\(k_\varphi (N m/rad)\) & 189506&
\(c_\varphi (N m / rad s) \) & 6364 \\
\(r_{eff} (m)\) & 0.393 &
\(I_w (kg m^2)\) & 1.1\\
\(\mu_x\) & 1 &
\(\mu_y\) & 1 \\
\(Q_{max} (N m)\) & 1600 &
\(Q_{min}(N m)\) & -1600 \\
\(\delta_{max} (rad) \) & 1&
\(RI_c \) & 0.7 \\

\(C_{\sigma} (N)\) & 80000&
\(C_{\alpha} (N/rad)\) & 47275\\ 
\(k_{usd} (rad s /m^2)\) & 0.4\\
\hline
\end{tabular} 

\end{center}
\end{table}

The MPC parameters are given in Table \ref{tab:mpc_ev_parameters}. Additionally, it is assumed that at the start of the horizon the weight is evenly distributed at the four vehicle corners. Moreover, the lateral forces on the tires are zero (i.e, the vehicle is moving in a straight line). Initially, the torques on the rear wheels are \( 100 N m\) and the other actuators aren't used. Ultimately, the MPC would use torque vectoring to ensure the vehicle is laterally stable while tracking the driver's desired yaw rate.

\begin{table}[htbp]
\caption{MPC Parameters for General Electric Vehicle.}
\label{tab:mpc_ev_parameters}
\begin{center}
\begin{tabular}{l|l|l|l}
\hline
Parameter & Value &Parameter&Value\\
\hline
\(N\) & 10 & \(M\) & 250\\
\(u (kph)\) &80 & \(T_s (s)\) & 0.1 \\ 
\(x_0 (SI)\) & \(
\begin{bmatrix}
    0,0,0,0,53,53,53,53
\end{bmatrix}^T\) 
& \(\sigma\) & 0.1\\
\(\alpha_{max} (deg)\) & 6 & \(\delta_d (rad)\) &0.15 \\
\hline
\end{tabular}
\end{center}
\end{table}

\subsection{VHS Vehicle Modal Parameters for Dallara AV-21} \label{sec:parameters-dallara}
In this section, we outline our specific model implementation for the Dallara AV-21 chassis. We have good estimates of most of the parameters defined in Table \ref{tab:av21_parameters} based on the real DAllara AV-21 chassis, which are then used in our later simulations in Section \ref{sec:experiment}. However, some potential values of the parameters are estimated for sanity checks. For example, the friction coefficient values have been set to 1; but in reality, they would need to be estimated in real time. 
\begin{table}[htbp]
\caption{Vehicle Parameters for Dallara AV-2.}
\label{tab:av21_parameters}
\begin{center}
\begin{tabular}{l|l|l|l}
\hline
Parameter&Value&Parameter&Value\\
\hline
\(r_{eff} (m)\) & 0.29 & \(l_f (m) \) & 1.6566 \\
\(l_r (m) \) & 1.3152 & \(m (kg)\) & 803.182 \\
\(m_s (kg)\) & 672.2 & \(m_f (kg) \) & 355.45 \\
\(I_{xx} (kg m^2) \) & 200 & \(h_s (m)\) & 0.1\\
\(I_z (kg m^2) \) & 1200 & \(C_\sigma (N) \) & 10000 \\
\(C_\alpha (N/rad) \) & 8000 & \(\mu_x\) & 1\\
\(\mu_y \) & 1 & \(K_s (N/m)\) &10000\\
\(b_s (N/m s) \) & 1000 & \(\delta_{max} (rad) \) & 0.1\\
\hline
\end{tabular}
\end{center}
\end{table}

These values can be gathered from any road or track racing vehicles. The value of the effective tire radius can be determined based on the tire radius ($r_w$) and the static tire radius ($r_{stat}$: the distance from the wheel hub to the point of contact with the floor) as:
\begin{align}
    r_{eff} = \frac{\sin(\cos^{-1}(\frac{r_{stat}}{r_w}))*r_w}{\cos^{-1}(\frac{r_{stat}}{r_w})}.
\end{align}

The quantities $l_r$ and $l_f$ are relatively intuitive to gather as the distance from the center of gravity of the vehicle to the rear and front axles respectively. From these and the total vehicle mass, the quantities $m_f$ and $m_r$ are determined as
\begin{align}
    m_f = m \frac{l_r}{l},\\
    m_r = m\frac{l_f}{l}.
\end{align}

Furthermore, $m_s$ is measured as the mass of the car without the suspension system and the tires and rims. The quantity $h_s$ has been determined through measuring up the suspension system and using the software simulator \emph{ChassisSim} \footnote{\url{https://www.chassissim.com/}.} to determine the roll center location. Then, we subtract that from the center of gravity height to get $h_S$. The tire stiffness coefficients were determined from historical tire data and the suspension parameters from historical suspension data. Moreover, the optimum method of computing the coefficient of friction is through the use of a Kalman Filter as in \cite{lastpaper}. Finally, there are various experiments to determine the moments of inertia, although some of them may be excessively complicated. One method of doing this is by approximating the vehicle as a rectangular slab and computing the moments of inertia in each direction accordingly.

The values of the various parameters in Table III have been attained either through measurement, as in the case of the various lengths, or from the team's historically gathered data.

\section{Experiment} \label{sec:experiment}
\subsection{General Electric Vehicle Model}
\begin{figure}[h!]
\centering
\includegraphics[width=80mm]{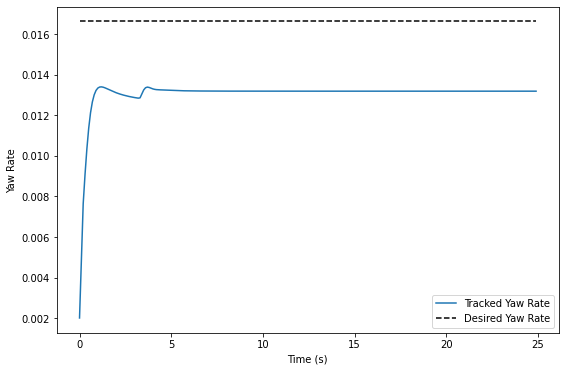}
\caption{Yaw rates of the General EV model.}
\label{figure: generalevyawrate}
\end{figure}

\begin{figure}[h!]
\centering
\includegraphics[width=80mm]{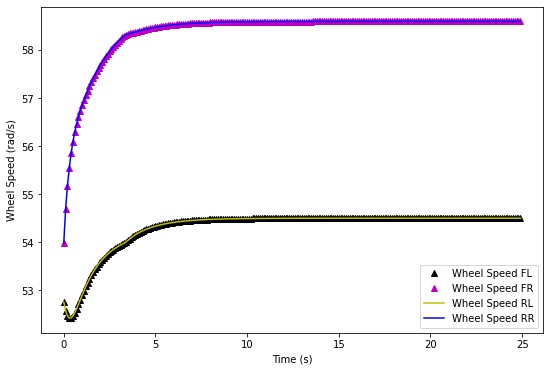}
\caption{Wheel speed of the General EV model.}
\label{figure: generalevwheelspeed}
\end{figure}

Under the conditions shown in Table \ref{tab:mpc_ev_parameters}, our results for the simulation of the response of an ideal vehicle with our MPC controller implemented is shown in Figures \ref{figure: generalevyawrate} and \ref{figure: generalevwheelspeed}. The actuator configuration matrix is selected such that the vehicle is controlled purely by torque vectoring. We can see clearly that the steady state yaw rate is different from the commanded one by the driver. The reason for this is that this is the maximum attainable yaw rate that ensures the constraints are satisfied and the vehicle is stable, i.e., the driver is safe. Moreover, Figure \ref{figure: generalevwheelspeed} shows the differential between the left and right wheel speeds, which is the selected method of control for turning the vehicle.

\begin{table}[htbp]
\caption{MPC Parameters for Dallara AV-21.}
\label{tab:av21_mpc_parameters}
\begin{center}
\begin{tabular}{l|l|l|l}
\hline
Parameter & Value &Parameter&Value\\
\hline
\(N\) & 50 & \(M\) & 140\\
\(u (kph)\) &180 & \(T_s (s)\) & 0.05 \\ 
\(x_0 (SI)\) & \(
\begin{bmatrix}
    0,0,0,0,0
\end{bmatrix}^T\) 
&\(\alpha_{max} (deg)\) & 6  \\
\hline
\end{tabular}
\end{center}
\end{table}

\subsection{Simulation for Dallara AV-21 Racing}
\begin{figure}[h!]
\centering
\includegraphics[width=90mm]{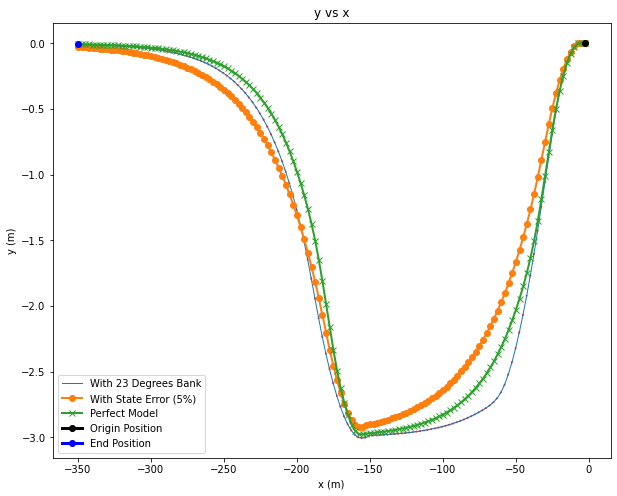}
\caption{Simulated trajectories of Dallara AV-21.}
\label{figure: dallaralfs}
\end{figure}

\begin{figure}[h!]
\centering
\includegraphics[width=43mm]{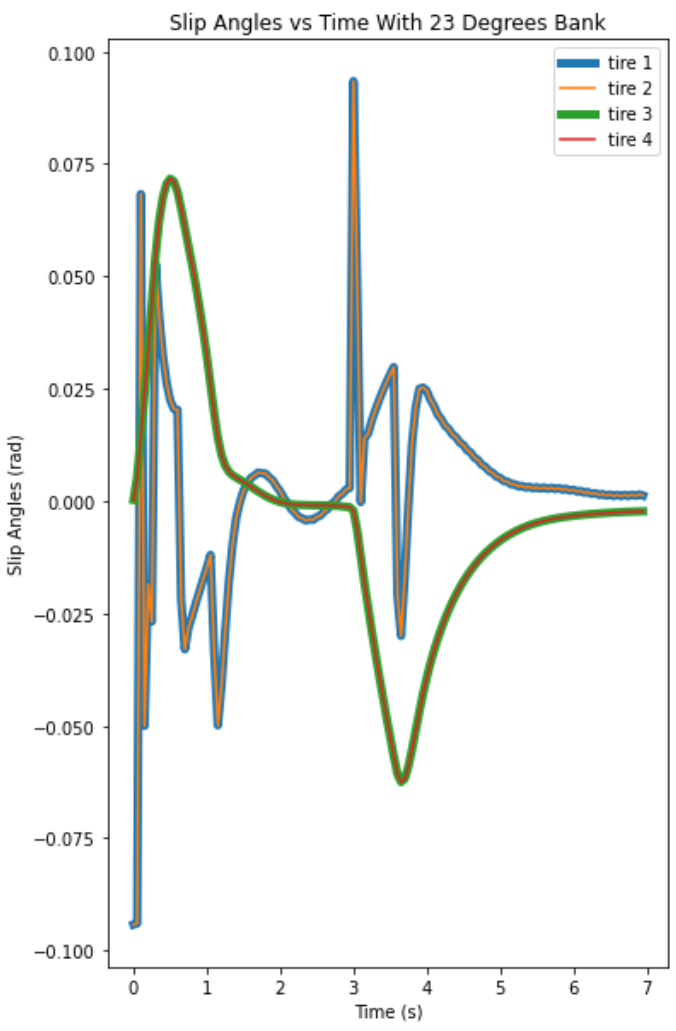}
\includegraphics[width=43mm]{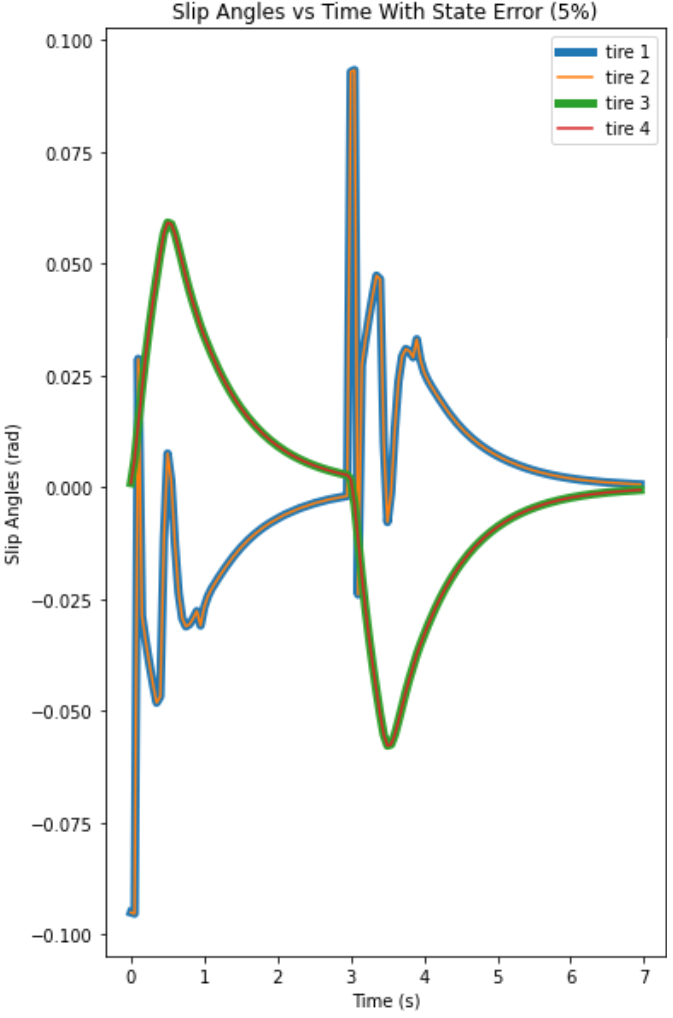}
\includegraphics[width=43mm]{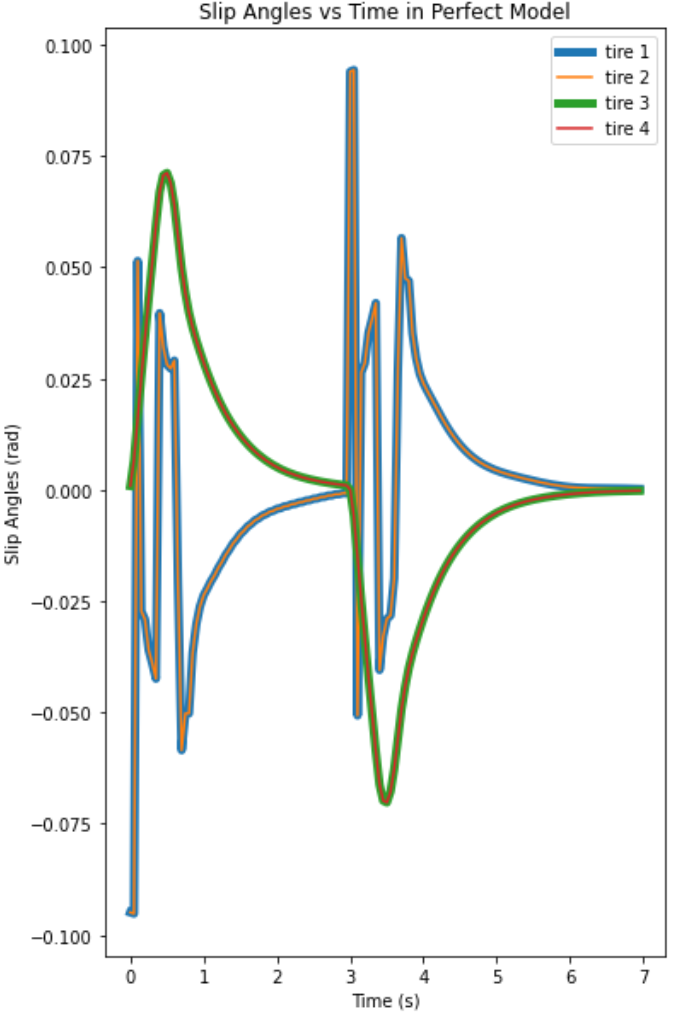}
\caption{Slip angles of Dallara AV-21 in three cases.}
\label{figure: dallaralfs-sa}
\end{figure}

\begin{figure}[h!]
\centering
\includegraphics[width=43mm]{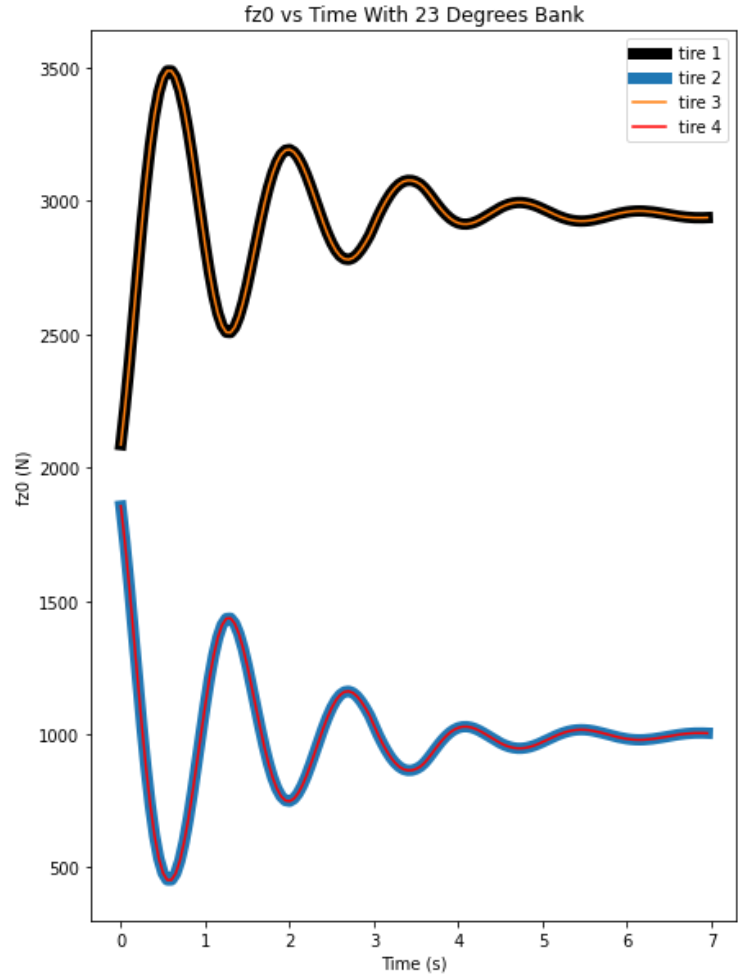}
\includegraphics[width=43mm]{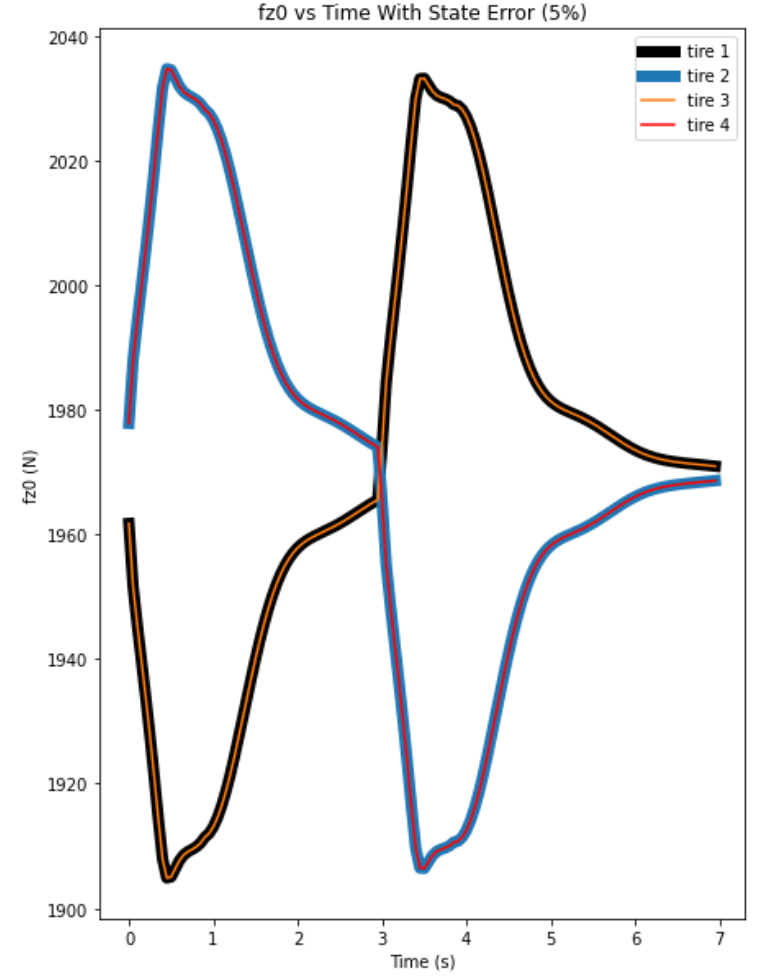}
\includegraphics[width=43mm]{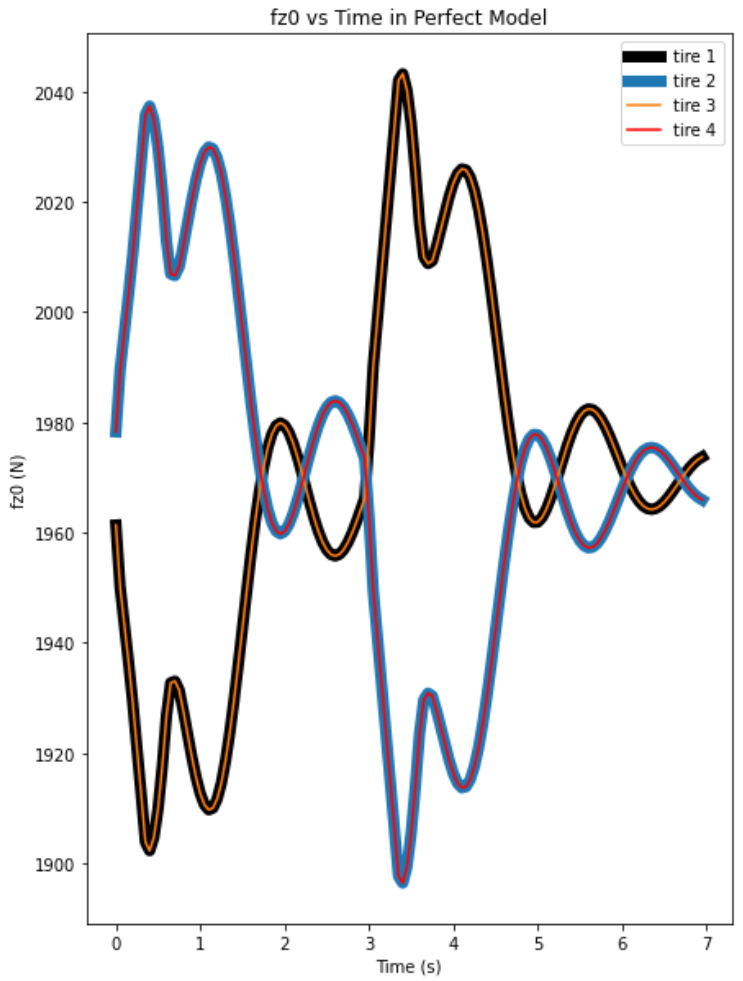}
\caption{Tire loads of Dallara AV-21 in three cases.}
\label{figure: dallaralfs-fz}
\end{figure}

The simulation run for the Dallara was a little different, resembling an overtaking maneuver. Due to the autonomous nature of this vehicle, the input is supplied in terms of checkpoints. Specifically, at time 0, the vehicle is at the origin. The first checkpoint is at x = 150 m and y = -3 m. Then, the vehicle should slot back in front of the opponent vehicle at x = 350 m and y = 0. Moreover, to ensure robustness of our controller, we perform this test under three scenarios:
\begin{enumerate}
\item with perfect model knowledge;
\item with a banking angle present;
\item with a 5\% downward error in the predicted states by the model predictive controller
\end{enumerate}
The results of the three simulations are shown in Figures \ref{figure: dallaralfs}, \ref{figure: dallaralfs-sa} and \ref{figure: dallaralfs-fz}.

In Figure \ref{figure: dallaralfs}, we can observe that for all three cases, the maneuver is completed successfully. We further observe that model uncertainties slow down the MPC's performance, but the vehicle is still able to complete the overtake. Moreover, we see that with a 23 degree banking angle, the vehicle steers out much quicker, due to the additional stability provided by the presence of the banking angle.

In Figure \ref{figure: dallaralfs-sa}, we can clearly see that the vehicle slip angles are bounded to be within 0.1 radians for all 4 wheels, which is our criteria for ensuring stability. Finally, Figure \ref{figure: dallaralfs-fz} displays the load transfer between the left and right wheels during the overtake. We observe that as the vehicle is turning out, the right tires carry more load and vice versa as it turns in. This is exactly what we expect for a typical vehicle under steering. Also, in the presence of a high banking angle, the left wheels always carry more load that the right wheels, which is perfectly expected. The oscillations in the normal loads could be decreased by tuning the MPC to put more weight on roll angle and roll rate.

\section{High-Fidelity Simulator Testing}
After the testing for the above results using a low-fidelity simulator, our team was able to further gain access to Open Source Self Driving Car (OSSDC) Initiative's simulator that opened the door for multiple new testing opportunities.\footnote{Based on the LGSVL simulator: \url{https://www.svlsimulator.com/}.}  The first and most obvious opportunity was the ability to test with a higher fidelity vehicle model and understand the vehicle's performance in that case. The second was the ability to integrate the controller with the remainder of the software stack, most importantly path planning and localization. Finally, this new horizon allowed us to explore and visualize the car performance around a track of our choosing, for which we selected the Las Vegas Motor Speedway (LVMS).

For the purposes of high-fidelity simulation, we were able to gather more accurate measurements of the vehicle parameters, including but not limited to: roll center location, tire cornering stiffness, and suspension stiffness and damping. Some of these parameters were attained using ChassisSim, where we used suspension geometry and various other measurements off the vehicle to attain parameters such as roll center location.  The results of test runs at different speeds are shown in Figures \ref{figure: hfs30} - \ref{figure: hfs55} below. The vehicle starts at the blue circle shown in the figures and tracks the pitlane line. After that, the car shifts to tracking the track line and keeps doing that. We can observe that at speeds up to 30 m/s the error is in the range of 7 cm. As speeds increase to 45 m/s, the error goes up to a maximum of 30 cm. However, at speeds beyond 55 m/s, the error is within 1.5 m.
\begin{figure}[h!]
\centering
\includegraphics[width=85mm]{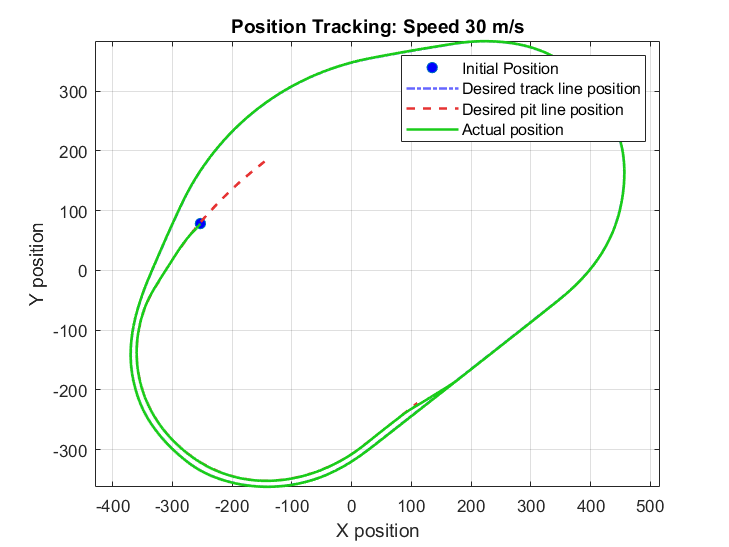}
\caption{Tracking path planner checkpoints at 30 mps.}
\label{figure: hfs30}
\end{figure} 

The reason behind this is fairly simple and could be easily resolved. Specifically, at speeds up to 45 m/s, we observe that the controller coded in Python runs at 50 Hz, which is the desired speed for our purpose. However, as the speed increases the feasible space decreases, resulting in slower solver speed, up to 30 Hz. As a result, the tracking error increases and at some point becomes too high (beyond 60 m/s). As a result, we believe that converting the controller to be written in C++ instead of in Python will alleviate the solver time issue and result in the proper tracking for VHS scenarios. 
\begin{figure}[h!]
\centering
\includegraphics[width=85mm]{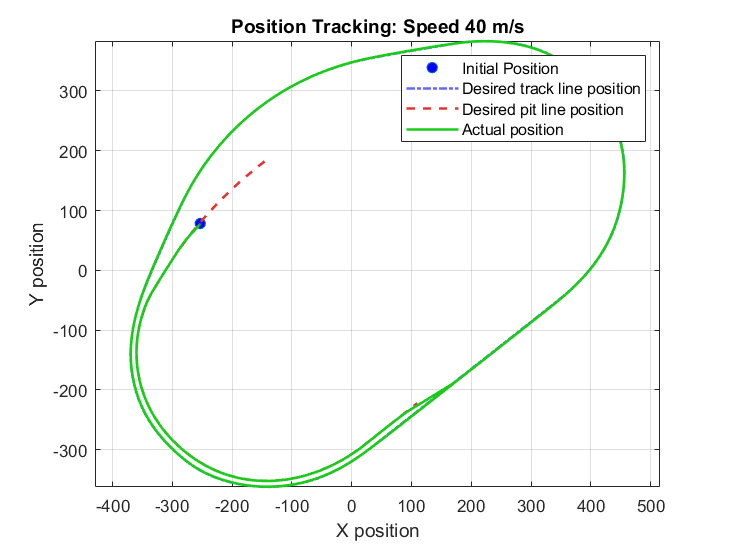}
\caption{Tracking path planner checkpoints at 40 m/s.}
\label{figure: hfs40}
\end{figure} 
\begin{figure}[h!]
\centering
\includegraphics[width=85mm]{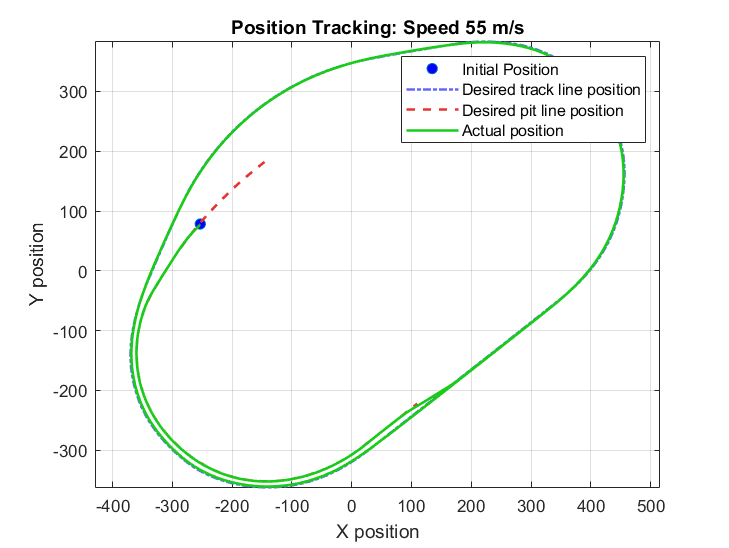}
\caption{Tracking path planner checkpoints at 55 m/s.}
\label{figure: hfs55}
\end{figure}\
\section{conclusion}
The work in this paper has resulted in a model predictive controller that accommodates roll dynamics to yield a better performance at the limits of stability. The linearization of the tire model resulted in a quicker solver time while being able to capture the changes in the tire force parameters resulting from a nonlinear model.

Ultimately, we can say that the MPC yields satisfactory results for both a general electric vehicle and the Dallara AV-21. Through various simulations, we are able to test the effects of varying different parameters, as well as study the robustness of the controller, under the stability constraint points mentioned. Future work includes converting the controller to C++ to alleviate the solver time issue, and consequently run the controller at higher speeds than those tested in this paper. This may impose tuning requirements that also fall under the umbrella of future work.

\end{document}